\documentclass[preprint,aps,pra]{revtex4}
\usepackage{times}
\usepackage{t1enc}
\usepackage[english]{babel}
\usepackage{graphicx}

\begin{document}

\title{Tunneling time calculations for general finite wavepackets
based on the presence time formalism}

\author{O. del Barco$^{1}$, M. Ortu\~no$^{1}$ and V.
Gasparian$^{2}$}

\affiliation{$^{1}$ Centro de Investigaci\'{o}n Optica y
Nanof\'{i}sica,
Universidad de Murcia, Spain \\
$^{2}$Department of Physics, California State University,
Bakersfield, USA}

\begin{abstract}
We analyze the tunneling time problem via the presence time
formalism. With this method we reproduce previous
results for very long wavepackets and we are able to calculate
the tunneling time for general wavepackets of arbitrary shape
and length.
The tunneling time for a general wavepacket is equal to the average
over the energy components of the standard phase time.
With this approach we can also calculate the time uncertainty.
We have checked that the results obtained
with this approach agree extremely well with numerical simulations
of the wavepacket evolution.
\end{abstract}

\maketitle

\section{Introduction}

The time spent by a particle in a given region of space is a
problem that has been approached from many different points of
view. There exits a huge literature on the tunneling phenomena for
electrons through a barrier. Landauer and Martin \cite{LM94}
pointed out that there is no clear consensus about simple
expressions for time in quantum mechanics (QM), where there is not
a Hermitian operator associated with it. Hartman \cite{HM62}
asserted that for opaque barriers the tunneling time is
independent of the barrier length and the time spent by the
particle in these regions can be less than the time required to
travel the same distance in vacuum. However, many physicists
hesitated to deal with Hartman´s results since a very fast
tunneling implies that the tunneling velocity or the average
velocity may become higher than the vacuum light velocity $c$. One
can define the traversal time as the time during which a
transmitted particle interacts with the region of interest
measured by some physical clock which can detect the particle's
presence after leaving the region. For electrons, this approach
can utilize the Larmor precession frequency of the spin produced
by a weak magnetic field acting within the barrier region
\cite{BZ671,BZ672,BT83}. Analogously, our group \cite{GO95}
proposed a clock based on the Faraday effect to measure the
interaction time for electromagnetic waves in a slab or a periodic
structure. Another approach consists of calculating the traversal
time of a particle through a barrier by following the behavior of
a wavepacket and determining the delay due to the structure of the
region. In this approach an emerging peak is not necessarily
related to the incident peak in a causative way \cite{LAN93}. The
phase time is the time which elapses between the peak of the
wavepacket entering the barrier and leaving it and can be defined
as the energy derivative of the phase.

Often, more than one tunneling time are involved in the problem.
One can define a time $\tau_y$ associated with the direction of
propagation and another time $\tau_z$ related to the transverse
direction. Büttiker \cite{BT83} assumed that the relevant
interaction time $\tau^{\rm BL}$ depends on both characteristic
times and is given by
\begin{equation}\label{timeBL}
\tau^{\rm BL} = \sqrt{\tau_y^2 + \tau_z^2},
\end{equation}
which is the so called B\"uttiker--Landauer time for transmitted
particles. Gasparian \emph{et al.} \cite{GP93,GOR95} introduced a
method based on Green's functions and obtained a complex tunneling
time, $\tau$. The real and imaginary part of this complex time
were related to $\tau_z$ and $\tau_y$, respectively.

Other method to calculate the traversal time uses the Feynman
path-integral approach of QM. Sokolovski and Baskin \cite{SB87}
applied this formulation to obtain a traversal time. Sokolovski
and Connor \cite{SC90} studied the tunneling time for wavepackets
via this path-integral approach. For the square potential barrier,
Fertig \cite{FT90}, using the path decomposition of Auerbach and
Kivelson \cite{AK85}, defined a propagator that corresponds to the
amplitude for tunneling between two points on opposite sides of
the barrier with initial energy $E$, summed over the Feynman paths
that spend a time $\tau$ inside the barrier.

In QM we can only measure quantities for which we have introduced
a Hermitian operator. For these quantities, expectation values can
be calculated and checked experimentally. However, time appears in
the standard quantum-mechanical approach only as a parameter and
therefore its expectation value is not defined. Moreover, Pauli
\cite{PAU58} argued that a self-adjoint time operator implies an
unbounded energy spectrum. In spite of this, many authors have
proposed time operators and others have developed formalisms for
arrival times in QM (for a review see Muga \emph{et al.}
\cite{MU00}). Allcock \cite{ALL69} was the first to focus on the
concept of arrival time, rather than on time operators, and
concluded that wave mechanics cannot give an exact and ideal
definition of arrival time. Werner \cite{WER86} overcame Pauli's
theorem by introducing non-self-adjoint operators in the framework
of positive operator valued measures (POVMs). Le\'{o}n \emph{et
al.} \cite{LE00} introduced a formalism for the calculation of the
time of arrival for particles travelling through a region with a
given potential energy. They employed quantum canonical
transformations from the free to the interacting cases to compute
the time of arrival in the context of the POVMs. However, it has
been criticized that this approach does not always recover the
classical expression for the arrival time when the effects of
non-commutativity of the operators involved may be neglected
\cite{BE01}.

Many authors have developed formalisms based on time operators
rather than arrival times in QM \cite{OR74,RE77,KO94,KO01}. In
these approaches, the average presence time at position $y$ for a
spatial wavepacket $\Psi(y,t)$ is defined as
\begin{equation}\label{pres}
\langle t (y) \rangle \equiv \frac{ \int_{-\infty}^{+\infty} \mathrm{d} t
|\Psi(y,t)|^2 \ t } { \int_{-\infty}^{+\infty} \mathrm{d} t
|\Psi(y,t)|^2},
\end{equation}
provided that this integral exists. Kobe \emph{et al.}\cite{KO94}
named the time operator whose average is given by Eq.\
(\ref{pres}) as the ''tempus'' operator, and one can study
efficiently the tunneling time through a barrier via the local
value of this operator \cite{KO01}.

We studied numerically the tunneling time for electronic
wavepackets in nanostructures and found that the finite size
effect of the incident wavepacket was relevant when treating
tunneling time in this spatial scale \cite{GO04}. The aim of this
work is to show first that the approach based on the presence time
gives equivalent results to standard treatments for very long
wavepackets. In second place, we want to study finite size effects
of the wavepacket in the tunneling time with this formalism, which
is specially suited for this problem. We also want to compare the
results obtained with the presence time formalism with those
obtained using the time of arrival approach of Le\'{o}n \emph{et
al.} \cite{LE00}.

The plan of the work is as follows. In Sec.\ II we introduce the
presence time formalism and apply this method to the simplest case
of free propagation. In Secs.\ III and IV we calculate, within the
framework of the presence time formalism, the tunneling time and
its uncertainty for a wavepacket which moves towards a rectangular
barrier, respectively. In Sec.\ V we present some numerical
results which include the finite size effect of the electronic
wavepacket in the tunneling time and its uncertainty for a
rectangular potential barrier. We also calculate the traversal
time and its uncertainty for photons crossing a set of layers with
a frequency gap. Finally, we summarize our results in Sec.\ VI.

\section{Presence time formalism}

The calculation of the average presence time can be performed in
terms of integrals over the energy, instead of integrals over time
as in Eq.\ (\ref{pres}). To this aim, it is convenient to consider
only scattering states incident with positive momenta so that
there is no energy degeneracy. Then we can define the energy
wavepacket in the following way
\begin{equation}
\Phi(y,E) = (2 \pi \hbar)^{-1/2} \int_{-\infty}^{\infty} dt \
\Psi(y,t) \ \exp[i (E t) / \hbar],
\end{equation}
where $\Psi(y,t)$ is the physical wavepacket in the space
representation. We can write the average presence time, Eq.\
(\ref{pres}), as a expectation value of the energy derivative
operator $-i \hbar
\partial_E$ in the energy representation \cite{EGUS00}
\begin{equation}\label{exptempus}
\langle t(y) \rangle = \frac{1}{P} \int_{0}^{\infty} \mathrm{d} E
\ \Phi^{\ast}(y,E) \left[- i \hbar \left( \partial\over\partial E
\right) \right] \Phi(y,E),
\end{equation}
where $P$ is the normalization factor
\begin{equation}
P(y) =\int_{0}^{\infty} \mathrm{d}
E \ |\Phi(y,E)|^2.
\end{equation}
We will asume that the energy wavefunctions $\Phi(y,E)$ are
continuous, differentiable and square integrable in the energy
variable. If we further restrict to functions satisfying
$\Phi(y,E=0)=0$ then the energy derivative operator $-i \hbar
\partial_E$ is Hermitian \cite{OR74}. For shortness, we
will refer to this operator as $\widehat{T}$ from now on.

To illustrate the presence time formalism, we consider first a
wavepacket propagating in free space. At $t=0$ this
wavepacket is peaked at $y_0$, has an spatial width $\Delta y$ and
moves to the right. The components of the wavepacket in the energy
representation are
\begin{equation}\label{wavfree}
\Phi(y,E) =  G(E) \ \exp\left[ i \ k (y - y_0) \right],
\end{equation}
where $G(E)$ is a normalized weight peaked at $E_0$ with an energy
width $\Delta E$, and $k(E)=\sqrt{2mE}/\hbar$ is the corresponding
wavenumber. Substituting expression (\ref{wavfree}) in Eq.\
(\ref{exptempus}) we obtain for the expectation value of
$\widehat{T}$ at a point $y$
\begin{equation}\label{expfree}
\langle \widehat{T} (y)\rangle = \frac{1}{P} \ \int_{0}^{\infty}
\mathrm{d} E \ G^2(E) \ \left[ \tau_{\rm cl}(y,E) - i \ \tau_{\rm G}(E)
\right],
\end{equation}
where $\tau_{\rm cl}(y,E)$ is the time it takes the particle to
travel from $y_0$ to $y$ with velocity $\sqrt{2mE} / m$
\begin{equation}
\tau_{\rm cl}(y,E) = \frac{m \ (y - y_0)}{\sqrt{2mE}},
\end{equation}
and $\tau_{\rm G}$ is the partial derivative of the natural
logarithm of the weight $G(E)$ with respect to the energy,
\begin{equation}\label{tg}
\tau_{\rm{G}}(E) = \hbar \ {\partial\ln G(E) \over\partial E}.
\end{equation}
The real part of the time is the average of the
classical time at $y$ for a particle with energy $E$ weighted by
the probability density in the energy representation.

We can easily prove the hermiticity of $\widehat{T}$ in this case
by showing that the imaginary part of $\langle \widehat{T}
\rangle$ cancels. Introducing Eq.\ (\ref{tg}) into Eq.\
(\ref{expfree}) we can write the imaginary part of the expectation
value of $\widehat{T}$ in the following way
\begin{equation}
\mathrm{Im} \left[ \langle \widehat{T}(y) \rangle \right] =
\int_{0}^{\infty} \mathrm{d} E \ G(E) \ {\partial G(E) \over
\partial E} = \int_{a}^{b} \mathrm{d} G \ G =
\left[\frac{G^2}{2}\right]_{a}^{b} = 0,
\end{equation}
where we have assumed that $G(E)$ tends asymptotically to $0$ in
the energy limits.

Let's calculate the uncertainty of $\widehat{T}$ for the free case.
The expectation value of the square of $\widehat{T}$ is equal to
\begin{equation}\label{expt2free}
\langle \widehat{T}^2 (y) \rangle
= \int_{0}^{\infty} \mathrm{d} E \ G^2(E) \ \{ [ \tau_{\rm cl}^2(y,E) -
\widetilde{\tau}_{\rm G}(E) - \tau_{\rm G}^2(E) ],
\end{equation}
where $\widetilde{\tau}_{\rm G}$
is the energy derivative of $\hbar \tau_{\rm G}$. So, the uncertainty
of $\widehat{T}$ is given by
\begin{equation}
\Delta \widehat{T}(y) = \sqrt{ \langle \tau_{\rm cl}^2(y,E)
\rangle - \langle \widetilde{\tau}_{\rm G}(E) \rangle - \langle
\tau_{\rm G}^2(E) \rangle - [ \langle \tau_{\rm cl}(y,E) \rangle
]^2 },
\end{equation}
where the averages represent the integrals over the energy
weighted by $G^2(E)$.

Now we restrict ourselves to a gaussian weight of width $\Delta E$
centered at $E_0$. For $\Delta E \ll E_0$ one can easily find that
$\langle \tau_{\rm cl}^2(y,E) \rangle
\simeq [ \langle \tau_{\rm cl}(y,E) \rangle ]^2$ and that $ -
\langle \widetilde{\tau}_{\rm G}(E) \rangle \simeq 2 \ \langle
\tau_{\rm G}^2(E) \rangle \simeq \hbar^2 / (2 \ \Delta E^2) $ so
the uncertainty of $\widehat{T}$ can be written as
\begin{equation}\label{uncfree}
\Delta \widehat{T} \simeq \frac{\hbar}{\sqrt{2} \ \Delta E}
\simeq \frac{\Delta y}{v_0},
\end{equation}
where $\Delta y$ is the spatial width of the free propagating
wavepacket and $v_0$ its group velocity. So, we have shown
explicitly that the energy-time uncertainty relation is satisfied
for the definition of $\widehat{T}$ in the free case.

\section{Tunneling time for a rectangular barrier}

\begin{figure}
\includegraphics[width=.8\textwidth]{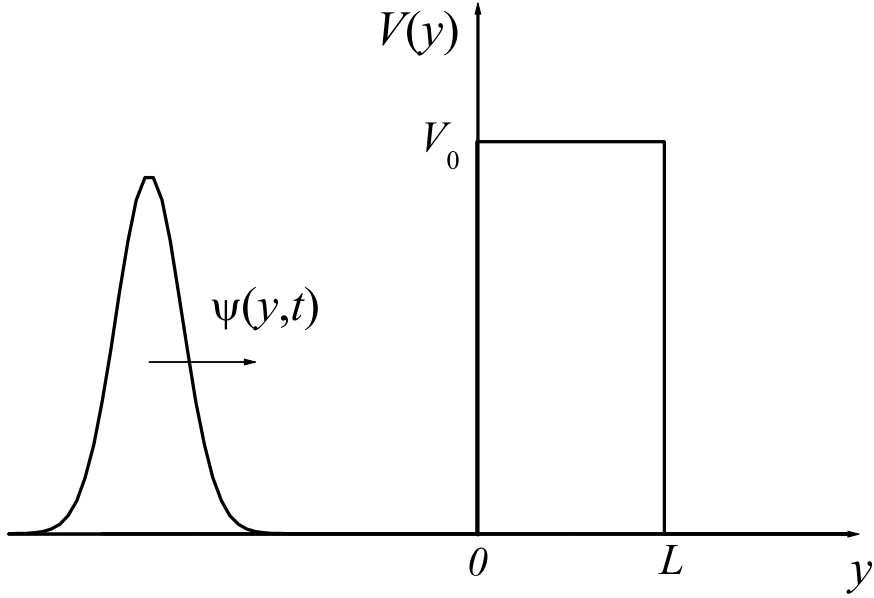}
\caption{A rectangular potential barrier placed between 0 and $L$
and a spatial wavepacket moving towards it.}\label{fig1}
\end{figure}

We now want to apply the presence time formalism to the tunneling
time problem. Let us consider a one-dimensional rectangular
potential barrier of height $V_0$ placed between $0$ and $L$ and a
spatial wavepacket $\Psi(y,t)$ which moves towards it (see Fig.\
\ref{fig1}). We calculate the expectation value of $\widehat{T}$
at $y=L$ with and without the barrier and, with our choice of
phases, the tunneling time $\tau$ will be equal to the difference
between these two times.

The components of the
wavepacket in the energy representation at the
right side of the barrier are given by
\begin{equation}\label{wavright}
\Phi_{\rm III}(y,E) =  G(E) \ |\widehat{t}(E)| \ \exp\left[ i ( k
(y -L)+ \varphi_{\rm t}(E) ) \right],
\end{equation}
where $|\widehat{t}(E)|$ is the modulus of the complex transmission
amplitude and $\varphi_{\rm t}(E)$ its phase. We have chosen the
phase in such a way that our origin of time is when the incident
wavepacket propagating freely would reach the left of the barrier,
but we have included the factor $ikL$ in the transmitted part so that
$\varphi_{\rm t}(E)$ does not accumulate
the phase for free propagation across the barrier.

To obtain the expectation value of the operator $\widehat{T}$ at
$L$, we first calculate the partial derivative of $\Phi_{\rm
III}(y,E)$ with respect to energy
\begin{eqnarray}\label{partright}
{\partial\Phi_{\rm III}(y,E)\over\partial E} &=& G(E) \left[
{\partial|\widehat{t}(E)|\over\partial E} + |\widehat{t}(E)| \
{\partial \ln G(E)\over\partial E} \right.
                \nonumber\\
& +&  \left. i \ |\widehat{t}(E)| \left( \frac{m \ (y
-L)}{\sqrt{2mE} \ \hbar} +
{\partial\varphi_{\rm{t}}(E)\over\partial E} \right) \right]
                \nonumber\\
&\times& \exp\{i \left[  k (y -L)+ \varphi_{\rm t}(E) \right]\}.
\end{eqnarray}
Multiplying Eq.\ (\ref{partright}) by $\Phi_{\rm III}^{\ast}(y,E)$
and integrating over the energy we can write the expectation value
of $\widehat{T}$ at $L$ as
\begin{equation}\label{expright}
\langle \widehat{T} (L)\rangle  = \frac{1}{P} \ \int_{0}^{V_0}
\mathrm{d} E \ |\Phi_{\rm III}(L,E)|^2 \  \tau_y(E),
\end{equation}
where $\tau_y(E)$ is the phase time
\begin{equation}\label{ty}
\tau_y(E) = \hbar \ {\partial\varphi_{\rm t}(E)\over\partial E}.
\end{equation}
$\tau_y(E)$, as given by Eq.\ (\ref{ty}), coincides with the
longitudinal characteristic time defined in the B\"uttiker
formalism \cite{BT83}. Let us remember that the condition for
hermiticity is that the weight $G(E)$ tends asymptotically to zero
when the energy tends to zero \cite{OR74}. One must ensure that
the weights considered satisfy this condition. Note that in Eq.\
(\ref{expright}) the integral over $E$ is up to $V_0$, since we
are assuming tunneling processes only.

Eq.\ (\ref{expright}) tells us that the tunneling time for a
general wavepacket of finite width is given by the average of
B\"uttiker time $\tau_y$ over the energy weighted by the
probability density in the energy representation at the right side
of the barrier, $|\Phi_{\rm III}(L,E)|^2$. Similar expressions can
be found in the literature (see, for example, Brouard \emph{et
al.} \cite{BS94} and Le\'{o}n \emph{et al.} \cite{LE00}). In both
cases, the time is written as an average of $\tau_y$ over the
momentum, instead of the energy. The only difference is in the
integration variable and, as we will see, it turns out to be very
small.

In order to obtain an analytical approximation for the tunneling time,
we assume a gaussian wavepacket of very small energy width $\Delta E$
and expand
$|\widehat{t}(E)|^2$ and $\tau_y(E)$ in Taylor series up to second
order near $E_0$. We arrive at the following result
\begin{equation}\label{exprightsec}
\langle \widehat{T}(L) \rangle  \simeq \tau_y(E_0) + \left(
\frac{1}{\hbar^2} \right) \left[ \tau_z(E_0) \
\widetilde{\tau}_y(E_0)  \right] (\Delta E)^2,
\end{equation}
where $\tau_z(E)$ is the tunneling time component related to the
transverse direction of propagation \cite{BT83}
\begin{equation}\label{tz}
\tau_z(E) = \hbar \ {\partial\ln |\widehat{t}(E)|\over\partial E},
\end{equation}
and $\widetilde{\tau}_y$ is the derivative of $\hbar \tau_y$ with
respect to energy. As we will see, Eq.\ (\ref{exprightsec}) is
valid up to values of spatial widths of the incident wavepacket
similar to the barrier length.

\section{Uncertainty of the tunneling time}

In this section we calculate the uncertainty of the tunneling time
through the barrier, which is equal to the sum of the
uncertainties of the incident and transmitted wavepackets. The
uncertainty of the tunneling wavepacket can be obtained through
the expectation value of the square of $\widehat{T}$ when the
system wavefunction is given by Eq.\ (\ref{wavright}). The
expectation value of $\widehat{T}^2$ at $L$ is
\begin{equation}\label{expt2right}
\langle \widehat{T}^2 (L)\rangle =\langle \tau_y^2 -
\widetilde{\tau}_z - \widetilde{\tau}_{\rm G} - (\tau_z +
\tau_{\rm G})^2 \rangle,
\end{equation}
where $\widetilde{\tau}_z$ is the derivative of $\hbar \tau_z$
with respect to the energy.
So, the uncertainty of $\widehat{T}$ at $L$ can be expressed as
\begin{equation}\label{uncright}
\Delta \widehat{T} (L)= \sqrt{\langle \tau_y ^2 \rangle
- \langle \widetilde{\tau}_z \rangle - \langle
\widetilde{\tau}_{\rm G} \rangle - \langle (\tau_z + \tau_{\rm
G})^2 \rangle  -\langle \tau_y \rangle ^2}.
\end{equation}
In the next section we will use this equation to calculate
numerically the
uncertainty in the tunneling time for general wavepackets.

In the limit of very long spatial wavepackets, very narrows in
energy, we can obtain an approximate analytical expression for the
uncertainty in the tunneling time. If we expand $\tau_y$,
$\tau_y^2$, $\tau_z$ and $\tau_z^2$ in Taylor series up to second
order in $E$ near $E_0$, and consider again a gaussian weight of
width $\Delta E$ centered at $E_0$, one can easily see that
$\langle \tau_y ^2 \rangle \simeq \langle \tau_y \rangle^2$ and
that $2 \ \langle (\tau_z + \tau_{\rm G})^2 \rangle \simeq \ -
\langle \widetilde{\tau}_{\rm G} \rangle$. Neglecting terms
proportional to $\widetilde{\tau}_z$ we can write Eq.\
(\ref{uncright}) in the following way
\begin{equation}\label{uncrel}
\Delta \widehat{T} (L) \simeq \frac{\hbar}{\sqrt{2} \ \Delta E}
\simeq \frac{\Delta y}{v_0},
\end{equation}
where $\Delta y$ is the spatial width of the transmitted
wavepacket and $v_0$ the group velocity of the incident one. We
can see that this uncertainty is proportional to the spatial width
of the wavepacket and satisfies the energy-time uncertainty
relation. Eq.\ (\ref{uncrel}) is valid only when $\Delta y$ is
larger than the barrier width as we will show in the next section.

The uncertainty associated to the incident wavepacket is equal to
its spatial width divided by the group velocity, $\Delta y / v_0$.
So, the uncertainty in the tunneling time $\Delta \tau$ is the sum
of this uncertainty of the incident packet and Eq.\
(\ref{uncright}).

\section{Numerical results}

In this section we present numerical results about finite size
effects in the tunneling time for electrons and photons. We
calculate the delay time of the particle by following the
behaviour of its wavepacket when crossing the square barrier as
before of height $V_0$ and placed between $0$ and $L$. We consider
a gaussian wavepacket in momentum space, centered at $p_0$ and of
width $\Delta p$, initially peaked at $y_0$ which moves towards a
potential barrier.

We follow the time evolution of the transmitted and the incident
wavepackets to measure the time it takes the particle to traverse
the potential barrier. We calculate the position of the centroid
and extrapolate its movement for the incident wave up to the
beginning of the barrier. We call $t_1$ the time when the incident
peak would reach the barrier assuming that there are not
perturbations due to the presence of the barrier. We also
calculate the centroid of the transmitted peak and extrapolate
back its movement to the right of the barrier $L$. The
corresponding time is called $t_2$. The tunneling time is then
defined as the difference between $\tau=t_2-t_1$. This approach is
the most adequate to include the finite size effects.

\begin{figure}
\includegraphics[width=.5\textwidth]{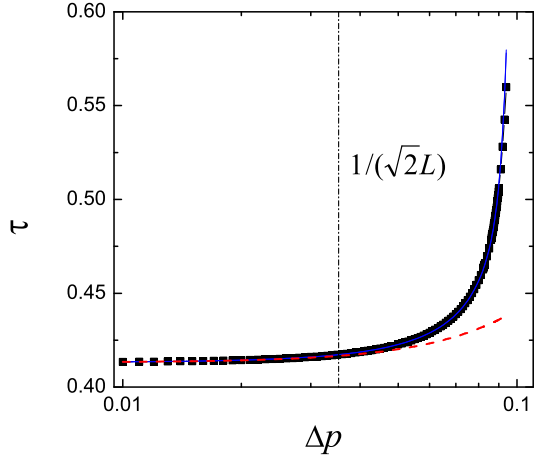}
\caption{Tunneling time $\tau$ versus the size of
the wavepacket in the momentum space, $\Delta p$, for an incident
electron with momentum $p_0 = 2.5$ and barrier parameters $V_0 =
5$ and $L = 20$. The numerical results are represented by squares,
the results based on the presence time formalism by the solid
curve and the second order approximation by the dashed
curve.}\label{graph1}
\end{figure}

In Fig.\ \ref{graph1} we represent the tunneling time, $\tau$,
versus the width of the incident wavepacket in the momentum
domain, $\Delta p$, for an incident electron with a momentum $p_0
= 2.5$. We use in all our work atomic units, i.e., $\hbar = m_e =
1$. The barrier parameters are $V_0 = 5$ and $L = 20$. The squares
represent the numerical results and the solid curve the results
obtained with the presence time formalism. We can see that this
curve fits very well the numerical results for all sizes of the
incident wavepacket. The dashed curve corresponds to the second
order approximation, Eq.\ (\ref{exprightsec}), and fits the
numerical results relatively well up to values of $1 / \Delta p$
of the order of the barrier length. For higher values of $\Delta
p$ the transmission coefficient cannot be replaced by a second
order approximation and more terms are needed to improve the
results. Our results based on the presence time formalism (solid
line in Fig.\ \ref{graph1}) basically coincide with the results
based on the approach of Le\'{o}n \emph{et al.} \cite{LE00}. The
former averages $\tau_y$ over the energy, while the latter
averages over momentum. For electrons, due to their non-linear
dispersion relation, both methods are not strictly equivalent, but
the difference between their results is always less than 0.5 \%\
in all cases studied.

For very small wavepackets in momentum space the results tend to
the real part of the time obtained with the Green function
approach \cite{GP93,GOR95}, which is the same as the $\tau_y$
component of the Buttiker time \cite{BT83}. The
Hartman effect is real, but it is not a paradox because only
occurs for very long wavepackets in real space so that the
uncertainty is much larger than the difference between the
tunneling time and the time it would take a free particle to cross
the barrier. We will study this problem more deeply in the context
of electromagnetic waves.

We now extend the previous calculations to the traversal time for
photons crossing a set of layers with a frequency gap. In this
case, we know that no signal can travel faster than light in
vacuum, so this is a good test of the possible constraints in
Hartman effect. For electromagnetic waves the expression of
$\widehat{T}$ is the same as for electrons but changing the energy
by the frequency divided by $\hbar$.

The incident electromagnetic wavepacket moves towards a periodic
arrangement of $N-1$ layers. Layers with index of refraction $n_1$
and thickness $d_1$ alternate with layers of index of refraction
$n_2$ and width $d_2$. For this periodic structure there exits a
frequency gap where evanescent modes can be found. The wavenumbers
in the layers of the first and second type are $k_1=\omega n_1/c$
and $k_2=\omega n_2/c$, respectively. Let us call $a$ the spatial
period, so $a=d_1+d_2$. The periodicity of the system allows us to
obtain analytically the transmission amplitude $t_{\rm N}$ using
the characteristic determinant method \cite{CU96}
\begin{equation}\label{trans}
t_{\rm N} = \exp(-i k_1 d_1) \times \left\{\cos(N\beta a/2)-i
\left( \frac{\sin{(N\beta a/2)}}{\sin (\beta a)} \right) \sqrt
{\sin^2(\beta a)+\left[ {\frac{k_1^2-k_2^2}{2k_1k_2}} \sin (k_2
d_2)\right]^2} \right\}^{-1},
\end{equation}
where $\beta$ plays the role of quasimomentum of the system, and
is defined by
\begin{equation}\label{beta}
\cos (\beta a) = \cos (k_1d_1)\cos (k_2 d_2) - \left(
{\frac{k_1^2+k_2^2}{2k_1k_2}} \right) \sin(k_1 d_1) \sin(k_2 d_2).
\end{equation}
When the modulus of the RHS of Eq.\ (\ref{beta}) is greater than
1, $\beta$ has to be taken as imaginary. This situation
corresponds to a forbidden frequency window. We perform a
simulation of the propagation of the wavepacket similar to that
for electrons.

\begin{figure}
\includegraphics[width=.5\textwidth]{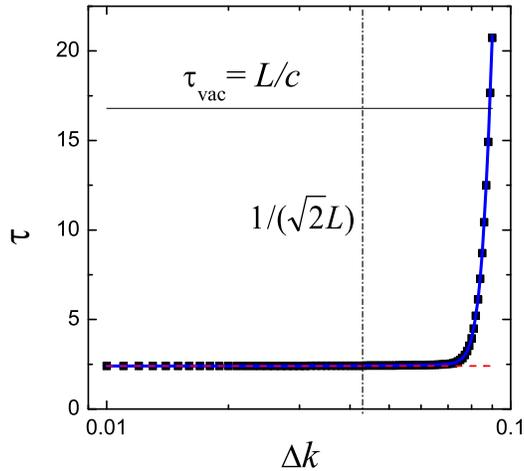}
\caption{Traversal time $\tau$ versus  $\Delta k$
for an electromagnetic wavepacket with $k_0 = 3.927$. The periodic
arrangement consists of $19$ layers with alternating indices of
refraction $2.0$ and $1.0$, and widths $0.6$ and $1.2$,
respectively. The squares represent the results obtained via the
wavepacket approach, the solid curve shows the results obtained
with the presence time formalism and the dashed curve the second
order approximation.}\label{graph3}
\end{figure}

In Fig.\ \ref{graph3} we represent the traversal time, $\tau$,
versus the width of the wavepacket in the wavenumber domain,
$\Delta k$, for an incident electromagnetic wavepacket with
momentum $k_0 = 3.927$, which corresponds to the centre of the
forbidden frequency window of our system. The units are set by the
choice $\hbar = c = 1$. The periodic arrangement consists of $19$
layers with alternating indices of refraction $2.0$ and $1.0$, and
widths $0.6$ and $1.2$, respectively, so the spatial length of the
structure is $L=16.8$. This periodic case satisfies the relation
$n_1 d_1 = n_2 d_2$ and most experimental setups use this periodic
arrangement \cite{ST93}. The numerical results are represented by
squares, while the solid curve corresponds to the results obtained
with the presence time formalism. We can see that this curve fits
very well the numerical results for all sizes of the incident
wavepacket. The dashed curve represents the second order
approximation and fits the numerical results relatively well up to
values of $1 / \sqrt{2} \Delta k$ of the order of the barrier
length. The limit of very narrow wavepackets in $\Delta k$ again
coincides with the real component of the time obtained with the
Green function approach. In the case analyzed, it is much smaller
than the crossing time of the structure at the vacuum speed of
light, represented by the horizontal line in  Fig.\ \ref{graph3}.
The crossing time remains smaller than the vacuum crossing time
for sizes of the wavepacket in real space up to half the width of
the structure.

\begin{figure}
\includegraphics[width=.5\textwidth]{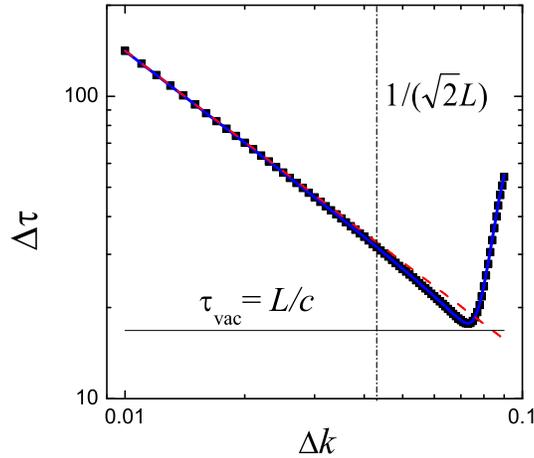}
\caption{Uncertainty of the tunneling time $\tau$
versus the size of the wavepacket in the wavenumber space, $\Delta
k$, for an incident electromagnetic wavepacket with central
wavenumber $k_0 = 3.927$. The periodic arrangement is the same as
in the previous example. The numerical results are shown by
squares and the solid curve represents the results obtained via
the presence time formalism. The dashed curve corresponds to the
second order approximation.}\label{graph4}
\end{figure}

In the circumstances analyzed in the previous paragraph, there is
no signal travelling faster than light at vacuum due to the large
uncertainty in the time. In Fig.\ \ref{graph4} we represent the
uncertainty of the tunneling time, $\Delta \tau$, versus the width
of incident the wavepacket in the wavenumber domain, $\Delta k$,
for the periodic structure of the previous example. The numerical
results are shown by squares and the solid curve represents the
results obtained via the presence time formalism. We can see that
the this curve fits very well the numerical results for all sizes
of the incident wavepacket. The dashed curve corresponds to the
second order approximation.

\section{Conclusions}

We have used the presence time formalism to calculate the
tunneling time and its uncertainty for finite size wave-packets.
In the simplest case of a one-dimensional rectangular potential
barrier the tunneling time is related to the expectation value of
the time operator $\widehat{T}$ at the right side of the barrier,
weighted by the transmitted wavepacket in the energy
representation, $|\Phi_{\rm III}(L,E)|^2$. This expectation value
is in turn given by an energy average of the phase time
$\tau_y(E)$.

For very long wavepackets the presence time formalism produces the
same results as previous approaches to the tunneling time problem
\cite{BT83,GO04}. For wavepackets of spatial size of the order of
the dimensions of the barrier, the results agree extremely well
with numerical simulations of wavepacket evolution. These results
are also in quite good agreement with our calculations based on
the time of arrival approach by Le\'{o}n \emph{et al.}
\cite{LE00}. Similar conclusions apply to the traversal time
problem of photons through dielectric structures in the frequency
gap region.

There is no fundamental problem with Hartman effect, because the
uncertainty in the time is larger than the advance in time with
respect to its vacuum value, whenever this difference is important
\cite{RO97}.
Our approach is particularly valuable for this type of problems,
since it is able to handle finite size effects of wavepackets.

\acknowledgements
The authors would like to acknowledge financial support from the
Fundacion Seneca and M.O. the Spanish DGI, project number BFM2003--04731.

\bigbreak

\end{document}